\documentclass[11pt, a4paper]{scrartcl}
\usepackage{lmodern}

\usepackage{amsmath, amssymb, amsthm, color, graphicx,
            subfigure, paralist} %
\usepackage[bookmarks,colorlinks,breaklinks]{hyperref}
\definecolor{dullmagenta}{rgb}{0.4,0,0.4}   
\definecolor{darkblue}{rgb}{0,0,0.4}
\definecolor{darkgreen}{rgb}{0,0.6,0}
\definecolor{darkred}{rgb}{0.6,0,0}
\hypersetup{linkcolor=darkblue,citecolor=blue,filecolor=dullmagenta,urlcolor=darkblue} 
\newcommand{\eqnumtag}{%
   \refstepcounter{equation}%
   \tag{\theequation}%
}

\newtheorem {theorem}{Theorem}
\newtheorem {lemma}[theorem]{Lemma}

\newtheorem {corollary}[theorem]{Corollary}

\newtheorem {defi}[theorem]{Definition}

\newtheorem {remark}[theorem]{Remark}
\newtheorem {remarks}[theorem]{Remarks}

\numberwithin{equation}{section}
\numberwithin{theorem}{section}
\numberwithin{figure}{section}

\usepackage{mathtools}

\newcommand{\beq}{\begin{equation}}
\newcommand{\eeq}{\end{equation}}
\newcommand{\Leq}[1]{\label{#1}\end{equation}}
\newcommand{\beqn}{\begin{eqnarray}}
\newcommand{\eeqn}{\end{eqnarray}}
\newcommand{\beqno}{\begin{eqnarray*}}
\newcommand{\eeqno}{\end{eqnarray*}}

\renewcommand {\l}{\left}

\newcommand {\ri}{\right}

\newcommand {\pa}{\partial}

\newcommand {\sign}{{\rm sign}}

\newcommand {\discr}{{\rm discr}}

\newcommand {\bR}{{\mathbb R}}
\newcommand {\bN}{{\mathbb N}}

\newcommand{\rstr}{{\upharpoonright}}

\newcommand{\cB}{{\mathcal B}} 
\newcommand{\cF}{{\mathcal F}} %

\newcommand{\cL}{{\mathcal L}}

\newcommand{\cR}{{\mathcal R}}

\newcommand{\bem}{\l(\! \begin{array}}
\newcommand{\eem}{\end{array}\!\ri)}
\newcommand{\bsm}{\left(\begin{smallmatrix}} 
\newcommand{\esm}{\end{smallmatrix}\right)}  

\newcommand{\qmbox}[1]{\quad\mbox{#1}\quad}

\newcommand{\s}{\sin}
\newcommand{\sh}{\sinh}
\newcommand{\co}{\cos}
\newcommand{\ch}{\cosh}

\newcommand{\rst}[1]{\ensuremath{{\mathbin\upharpoonright}%
\raise-.5ex\hbox{$\scriptstyle #1$}}}

\newcommand{\Mid}{\;\middle\vert\;}

\DeclareMathOperator{\arccosh}{arccosh}

\begin{document}
\title{The Problem of Two Fixed Centers: Bifurcation Diagram for Positive Energies}
\author{Marcello Seri\thanks{Department of Mathematics
University College London
Gower Street, London WC1E 6BT (UK), \texttt{m.seri@ucl.ac.uk}}
\date{}
}

\date{\today}

\maketitle

\begin{abstract}
  We give a comprehensive analysis of the Euler-Jacobi problem of
  motion in the field of two fixed centers with arbitrary relative
  strength and for positive values of the energy.  These systems
  represent nontrivial examples of integrable dynamics and are
  analysed from the point of view of the energy-momentum mapping from
  the phase space to the space of the integration constants.  In this
  setting we describe the structure of the scattering trajectories in
  phase space and derive an explicit description of the bifurcation
  diagram, i.e. the set of critical value of the energy-momentum map.
  
  
\end{abstract}
\tableofcontents\newpage
%

%
\section{Introduction}\label{I}
%

The study of two-centers Coulombic systems, both from a classical and
quantum point of view, is distributed along the last three centuries,
starting from pioneering works of Euler in 1760, Jacobi \cite{Jacobi}
in 1884 and Pauli \cite{Pauli} from a quantum mechanical point of view
in 1922.

Indeed, the models described by these systems are important both for
macroscopic and microscopic systems: in celestial mechanics they model
the motion of a test particle attracted by two fixed stars, and in
molecular physics they represent the simplest models for one-electron
diatomic molecules (e.g.  the ions ${\rm H}_2^+$ and ${\rm HHe}^{2+}$)
and appear as the first term in the Born-Oppenheimer approximation of
molecules.

Among the central features of this class of models is their
integrability and the separability in elliptical coordinates.  This
makes possible to introduce a number of significant reductions in the
study and therefore makes te model very suitable as a test field for a
number of questions.

Despite the age and the many properties of the problem, the research
around it goes up to the present days \cite{WDR, Kn, RecentSpecHydro}
and there are present many challenges that have to be addressed.

This work is a necessary step to prepare the foundations for the study
of the quantum resonances of the planar quantum mechanical two-centers
Coulomb system in the semi-classical limit \cite{SKDE14}.
Additionally the planar restriction of the two-centers problem arise
naturally in the analysis of the three-dimensional system one as an
essential prototypical building block \cite{thesis}.

A comprehensive analysis for the negative energies picture has been
given by Waalkens, Dullin and Richter \cite{WDR}. The classical
scattering was studied and described by Knauf and Klein \cite{KK, Kn}.
The contribution of this work is the completion of the phase space
picture and of the bifurcation diagram for the two-centers systems
with arbitrary relative strengths in the case of positive energies
(see Theorem~\ref{thm:defbifdia} and its corollaries). For positive
energies we are in a scattering situation and the orbits foliate the
space into families of diffeomorphic cylinders, the
bifurcation diagrams let us describe the different families of
trajectories and their properties.

Quantum resonances are a key notion of quantum physics: roughly
speaking they are scattering states (i.e. states of the essential
spectrum) that for long time behave like bound states
(i.e. eigenfunctions). They are usually defined as poles of a
meromorphic function, but there is not really a unique way to study
and define them \cite{zwo}. On the other hand, it is known that their
many definitions coincide in some settings \cite{hemartinez} and that
their existence is related to the presence of some classical orbits
``trapped'' by the potential.

The importance of this work in the setting of quantum resonances lies
in the strong connection between these and the structure of the
underlying classical system.  In fact it has been proven that there
are resonances generated by classical bounded trajectories around
local minima of the potential \cite{H-S} and that there are resonances
generated by closed hyperbolic trajectories or by non-degenerate
maxima of the potential \cite{BCD2, BCD3, Sj, GeSj}. The main
difference being in their asymptotic distance from the real axis in
terms of the semiclassical parameter $h$.

Even the presence or absence of the resonances is strictly related to
the classical picture. In fact it is possible to use some classical
estimates, called non-trapping conditions, to prove the existence of
resonance free regions (see for example \cite{BCD1,mart1,mart1b}).

The failure of the non-trapping condition for the two-centers problem
was already known in the literature \cite{CJK} as well as the presence
of a close hyperbolic trajectory for positive energies \cite{KK}. With
the present analysis we are able to explicitly identify the energies
associated with this hyperbolic trajectory and to find a positive
measure of positive energies associated to families of bounded
trajectories.  Their presence makes the present models a very good
candidate for the study of quantum resonances in presence of singular
potential \cite{SKDE14, thesis}.

{\bf Notation.}  In this article $0\not\in\bN$,
$\bR^{*}:=\bR\setminus\{0\}$ and $\bR_{+}:=(0,\infty)$.

\section{The classical problem of two Coulomb centers}
\label{ch:classic}

We consider the classical Hamiltonian function on the cotangent bundle
$T^{*}Q_2$ of $Q_{2} := \bR^{2}\setminus\{s_1,s_2\}$ relative to the
$2$-center potential given by:
\begin{equation}\label{1}
  H:T^*Q_2\to \bR\qmbox{,}
  H(p,q) := \frac{|p|^{2}}2 + \frac{-Z_1}{|q-s_1|} + \frac{-Z_2}{|q-s_2|}.
\end{equation}
This describes the motion of a test particle in the field of two
bodies of relative strengths $Z_i\in\bR^{*} =\bR\setminus\{0\}$, fixed
at positions $s_1\neq s_2\in\bR^2$.
By the unitary realisation $Uf(x) := |\det A|^{-1/2} f(Ax +b)$ of an
affinity of $\bR^2$ we assume that the two centers are at $s_1 := a :=
\bsm 1\\0\esm$ and $s_2:=-a$.

\subsection{Elliptic coordinates}\label{sec:pec3dintro}
The restriction to the rectangle $M:=\bR_{+}\times(-\pi,\pi)$ of the
map
\beq
  G:\bR^2\to\bR^2\qmbox{,} \bsm\xi \\
  \eta \esm \mapsto \bsm \ch(\xi)\co(\eta) \\
  \sh(\xi) \s(\eta) \esm
\Leq{elliptic}
defines a $C^{\infty}$ diffeomorphism \beq G:M\to G(M) \Leq{def-G}
whose image $G(M)=\bR^{2}\setminus(\bR\times\{0\})$ is dense in
$\bR^{2}$. Moreover it defines a change of coordinates from
$q\in\bR^{2}$ to $(\xi,\eta)\in M$. These new coordinates are called
\emph{elliptic coordinates}.

\begin{figure}[h!]
\begin{center}
 \includegraphics[width=0.4\linewidth]{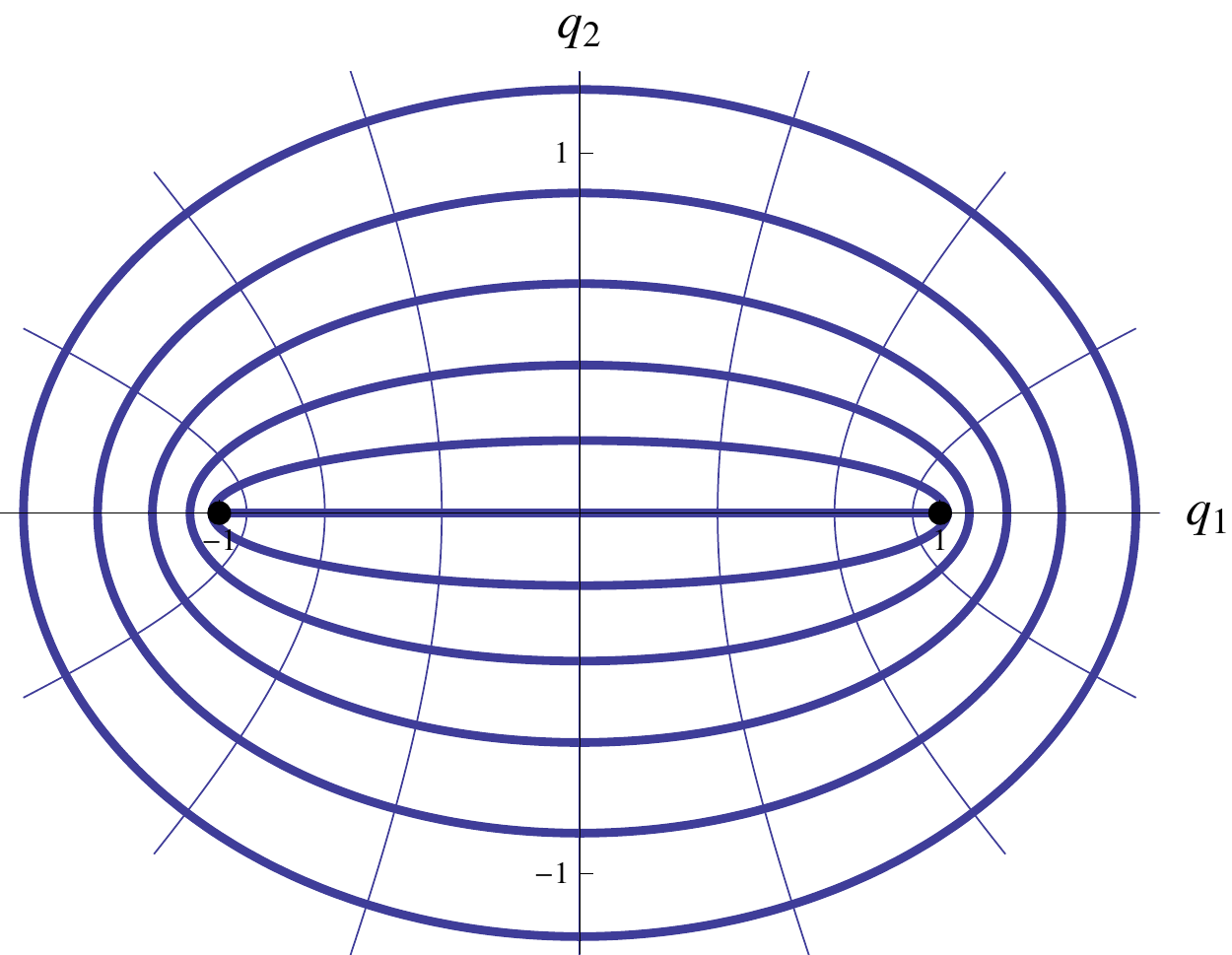}
\caption{\small Elliptic coordinates.}
\label{fig:coordPE}
\end{center}
\end{figure}

\begin{remarks}\label{rmk:coord}
\begin{compactenum}
\item In the $(q_{1},q_{2})$-plane the curves $\xi=c$ are ellipses
  with foci at $\pm a$, while the curves $\eta=c$ are confocal half
  hyperbolas, see Figure \ref{fig:coordPE}.
\item The Jacobian determinant of $G$ equals
  \beq
    F(\xi,\eta) := \det(DG(\xi,\eta)) =
    \sh^{2}(\xi)+\s^{2}(\eta) =
    \ch^2(\xi)-\co^2(\eta).
  \Leq{def-F}

  Thus the coordinate change \eqref{elliptic} is degenerate at the
  points $(\xi,\eta)\in\{0\}\times\{0,\pm\pi\}$ in $\overline{M}$.
  For $\xi=0$ the $\eta$ coordinate parametrizes the $q_{1}$-axis
  interval between the two centers. For $\eta=0$ ($\eta=\pm\pi$) the
  $\xi$ coordinate parametrizes the positive (negative) $q_{1}$-axis
  with $|q_{1}|>1$.  \hfill$\Diamond$
\end{compactenum}
\end{remarks}

\subsection{Hamiltonian setting}

\begin{lemma}
  Using $G$ defined in \eqref{def-G}, and $Z_\pm := Z_2 \pm Z_1$, $H$
  is transformed by the elliptic coordinates into
  \beq
    H\circ {(G^{-1})^*}(p_\xi,p_\eta,\xi,\eta) =
    \frac1{F(\xi,\eta)}\big(H_{1}(p_\xi,\xi) + H_{2}(p_\eta,\eta)\big)
  \Leq{eq:sep-midstep0}
  where $(G^{-1})^*:T^*M\to T^*Q_2$ is the cotangential lift of
  $G^{-1}$, and
  \beq
    H_{1}(p_\xi,\xi) := \frac{p_{\xi}^{2}}2 - Z_{+} \ch(\xi)\qmbox{,}
    H_{2}(p_\eta,\eta) := \frac{p_{\eta}^{2}}2 + Z_{-} \co(\eta).
  \Leq{eq:sep-midstep1}
  There are two functionally independent constants of motion $H$ and
  $L:=H_{1} - \ch^{2}(\xi)H$ with values $E$ and $K$ respectively.
\end{lemma}

\proof Although the lemma is well-known (see, {\em e.g.}, Thirring
\cite{Thi}, Sect.\ 4.3), we indicate its proof, in order to introduce
some notation: If we apply to $H$ the canonical point transformation
induced by the elliptic coordinates \eqref{def-G}, the potential is
transformed as
\beq
  V\circ G(\xi,\eta) =
  - \frac{Z_1}{ |q-a| } - \frac{Z_2}{ |q+a| }
  = {-\frac{Z_+\ch(\xi)-Z_-\co(\eta)}{F(\xi,\eta)}},
\Leq{eq:potTransf0}
and the momenta $p = (p_1, p_2)$ are transformed according to
\[
\bsm
p_1 \\
p_2
\esm
= (DG(\xi,\eta)^{-1})^t
\bsm
p_\xi \\
p_\eta
\esm.
\]
From $(DG(\xi,\eta)^{-1})(DG(\xi,\eta)^{-1})^t =
(DG(\xi,\eta)^{t}DG(\xi,\eta))^{-1}$ and
\[
(DG(\xi,\eta)^{t}DG(\xi,\eta)) =
\bsm
F(\xi,\eta) & 0 \\
0 & F(\xi,\eta)
\esm
\]
we get that the Hamiltonian is transformed into
\eqref{eq:sep-midstep0}.
\qed\\[1mm]

Given an initial condition $x_0\in T^*Q_2$ we set $E:=H(x_0)$.

Equation \eqref{eq:sep-midstep0} can be again separated \cite[Lemma
10.38]{Kn11} moving to the extended phase space and using a new time
parameter $s$ defined by
\[
\frac{dt}{ds} = F(\xi,\eta).
\]
One obtains the new Hamiltonian
\[
\widetilde H:=F(\xi,\eta)(H-E) = H_\xi + H_\eta
\]
where
\beq
  H_\xi(p_\xi,\xi) :=
  H_{1} - \ch^{2}(\xi)E \qmbox{and} H_\eta(p_{\eta},\eta)
  := H_{2} + \co^{2}(\eta)E.
\Leq{eq:hxihaetasep0}
On the submanifold $\widetilde H^{-1}(0)$, $\widetilde H$ describes
the time evolution of $H^{-1}(E)$ up to a time
reparametrisation. Therefore we have a second constant of motion other
than $H$:
\beq
  L := H_{1} - \ch^{2}(\xi)E =
  - \l( H_{2} + \co^{2}(\eta)E \ri).
\Leq{eq:sep-thirdconst}

Setting $K := H_{\xi}(x_{0}) = -H_{\eta}(x_{0})$ we have two constants
of motion $H$ and $H_{\xi}$ whose values are denoted respectively $E$
and $K$.  Notice that these functions are generally independent in the
following sense. Being real analytic functions, the subset of phase
space where independence is violated, is of Lebesgue measure zero.
\qed

\begin{remarks}\label{rem:setting}
\begin{enumerate}
\item By the previous proof, we can restrict our attention to the
  phase space $T^*Q_2$ with Hamiltonian
  \[
    \widehat H (p_{\xi},p_{\eta},\xi,\eta) = 
    H_\xi(p_{\xi},\xi) +
    H_\eta(p_{\eta},\eta),
  \]
  where the $H_\xi$ and $H_\eta$ are defined in
  \eqref{eq:hxihaetasep0} and have the form
  \beq
    H_{\xi}(p_{\xi},\xi) := \frac{p_{\xi}^{2}}2 + V_{\xi}(\xi),
    \qquad
    H_{\eta}(p_{\eta}, \eta) := \frac{p_{\eta}^{2}}2 + V_{\eta}(\eta).
  \Leq{eq:effPot}
  Here $V_{\xi}$ and $V_{\eta}$ are defined by
  \beq
    V_{\xi}(\xi) := - Z_+\ch(\xi) - E\ch^2(\xi)
    \qmbox{,}
    V_{\eta}(\eta):= Z_-\co(\eta) + E\co^2(\eta).
  \Leq{eq:V1V2Classical}

\item Notice that the trajectories may cross the $q_1$-axis (where the
  prolate elliptic coordinate are singular) and even collide with the
  two centers at $\pm a=(\pm1,0)$. There are some different ways to
  regularise the motion both in the planar and spatial cases (and with
  an arbitrary number of centers), we refer the reader to
  \cite[Chapter 4-5]{Kn}, \cite[Chapter 3]{KK} and \cite[Remark
  11.24]{Kn11} for more details and references. \label{rem:regularity}
  \hfill$\Diamond$
\end{enumerate}\end{remarks}

\section{Bifurcation diagrams}\label{sec:BifDia}

Taken together, the constants of motion define a vector valued
function on the phase space of a Hamiltonian.  We can study the
structure of the preimages of this function (its level sets), in
particular their topology. In the simplest case the level sets are
mutually diffeomorphic manifolds.

\begin{defi} {\bf (see \cite[Section 4.5]{AM})}\label{def:bs}
  Given two manifolds $M, N$, $f\in C^{\infty}(M,N)$ is called {\bf
    locally trivial} at $y_{0}\in N$ if there exists a neighbourhood
  $V\subseteq N$ of $y_{0}$ such that $f^{-1}(y)$ is a smooth
  submanifold of $M$ for all $y\in V$ and there there is a map $g\in
  C^{\infty}(f^{-1}(V),f^{-1}(y_{0}))$ such that $f\times
  g:f^{-1}(V)\to V\times f^{-1}(y_{0})$ is a diffeomorphism.

  The {\bf bifurcation set} of $f$ is the set
  \[
   \cB(f) := \{ y_{0}\in N \mid f
   \mbox{ is not locally trivial at } y_{0}\}.
  \]
\end{defi}
Notice that if $f$ is locally trivial, the restriction
$g\rst{f^{-1}(y)}:f^{-1}(y)\to f^{-1}(y_{0})$ is a diffeomorphism for
every $y\in V$.
\begin{remark}
  The critical points of $f$ lie in $\cB(f)$ (see
  \cite[Prop. 4.5.1]{AM}), but the converse is true only in the case
  $f$ is proper (i.e. it has compact preimages).  \hfill$\Diamond$
\end{remark}

Define the function on the phase space as follows {(omitting a
  projection in the second component)}
\beq
  \cF := \bsm H \\ H_{\xi} \circ G^*\esm : T^{*}Q_2\to \bR^{2}.
\Leq{def:enmomMapping}
In what follows we characterise the bifurcation set $\cB(\cF)$.

\subsection{Bifurcations for planar motions}\label{sec:planarbif}
We have already discussed the regularisability of the problem in
Remark~\ref{rem:setting}.\ref{rem:regularity}. In what follows we
proceed similarly as \cite{WDR} but we consider the energy range $E\ge
0$.

It will be computationally useful to introduce a new coordinate
change. The restriction to $M^2:= (1,\infty)\times(-1,1)$ of the map
\[
  \bsm x \\ y \esm \in \overline{M^2}
  \mapsto
  \bsm \arccosh(x) \\ \arccos(y) \esm \in\bR^{2}
\]
defines a $C^\infty$ diffeomorphism
\beq
  G^2 : M^2 \to G^2(M^2)
\Leq{eq:def-G1}
with image $G^2(M^2) = \bR_+\times(0,\pi)\subsetneq M$. Therefore it
defines a change of coordinates from $(\xi,\eta)\in\bR_+\times(0,\pi)$
to $(x,y)\in M^2$.

\begin{lemma}\label{thm:hamsympl0}
  The diffeomorphism defined in \eqref{eq:def-G1} induces a
  symplectomorphism 
  $ \widehat {G}^2 : T^*(\bR_+\times(0,\pi)) \to
  T^*M^2. $
\end{lemma}
\noindent
{\it Proof.}
It is enough to choose the generating function
\[
S_2(\xi,\eta,p_x,p_y) :=
\big(\cosh(\xi), \cos(\eta)\big) \bsm p_x \\ p_y \esm.
\]
It induces a canonical transformation
\[
\widehat {G}^2 : (p_\xi,p_\eta,\xi, \eta) \mapsto (p_x,p_y,x,y)
\]
where
\[
\bsm x \\ y \esm
= \bsm
\frac{\pa S_2}{\pa p_x} \\ \frac{\pa S_2}{\pa p_y}
\esm = \bsm
\ch(\xi) \\
\co(\eta)
\esm,\;
\bsm p_\xi \\ p_\eta \esm
= \bsm
\frac{\pa S_2}{\pa \xi} \\ \frac{\pa S_2}{\pa \eta}
\esm = \bsm
\sh(\xi)p_x \\
\s(\eta)p_y
\esm\hfill\tag*{\qed}
\]

To cover the $(q_{1},q_{2})$-plane of the configuration space $Q_{2} =
\bR^{2}\setminus\{\pm a\}$ we need two half strips
$[1,\infty)\times[-1,1]$ (i.e. one for each sign of
$q_{2}$). Alternatively we can take the strip
\[
{ \bsm \xi\\\eta\esm  \in \widetilde{Q}_{2} :=} \bR\times [-\pi,\pi]
\]
as the modified configuration space: {For the map $ \bsm \xi\\
  \eta\esm \mapsto \bsm\ch(\xi) \\ \co(\eta) \esm$} it is a two-sheeted
cover with branch points at the foci. The two sheets are related by
the involution $I:(\xi,\eta)\mapsto(-\xi,-\eta)$ leaving the cartesian
coordinates $(q_{1},q_{2})$ unchanged. The symplectic lift of $I$ to
the phase space $T^{*}\widetilde{Q}_{2}$ equals
\[
  \widehat I: (p_{\xi},p_{\eta},\xi,\eta)
  \mapsto (-p_{\xi},-p_{\eta},-\xi,-\eta).
\]
Then $T^{*}Q_{2}$ is obtained from $T^{*}\widetilde{Q}_{2}$ by
factorisation with respect to ${\widehat I}$.

\begin{remark}\label{rem:boundsbif}
  An analysis of the extrema of $V_{\xi}$ and $V_{\eta}$ implies that
  the image $\cR$ of $(H,H_{\xi})$ in $\bR^{2}$ is bounded by the
  following curves. From $K = H_{\xi} \geq V_{\xi}$ we have $K\geq
  K_{+}(E)$ with \beq K_{+}(E) :=
\begin{cases}
- \infty, & E > 0 \\
-(Z_{+} + E), & E \leq \min\l(-\frac{Z_{+}}2, 0\ri) \\
\frac{Z_{+}^{2}}{4E}, & 0 \geq E >  \min\l(-\frac{Z_{+}}2, 0\ri)
\end{cases}
\Leq{eq:kplus}
and from $-K = H_{\eta} \geq V_{\eta}$ we have $K\leq K_{-}(E)$ with
\beq
K_{-}(E) :=
\begin{cases}
Z_{-} - E, & E \leq \frac{Z_{-}}2 \\
\frac{Z_{-}^{2}}{4E}, & E >  \frac{Z_{-}}2
\end{cases}.
\Leq{eq:kminus}
\end{remark}

The main objects of our analysis are transformed by the
symplectomorphism defined in Theorem \ref{thm:hamsympl0} as follows
\beq
\begin{array}{lclcl}
F(\xi,\eta) &\mapsto& \hat F(x,y)&:=& x^{2} - y^2, \\
V_{\xi}(\xi) &\mapsto&  V_{x}(x)&:=& -Z_{+}x -Ex^{2} , \\
V_{\eta}(\eta) &\mapsto& V_{y}(y) &:=&  Z_{-}y +Ey^{2}  , \\
H_{\xi}(p_{\xi},\xi) &\mapsto& H_{x}(p_{x},x)&:=&
  \frac{(x^{2}-1)p_{x}^{2}}{2} + V_{x}(x), \\
H_{\eta}(p_{\eta},\eta) &\mapsto& H_{y}(p_{y},y) &:=&
\frac{(1-y^{2})p_{y}^{2}}{2} + V_{y}(y).
\end{array}
\Leq{eq:renamedClassicalFcts}

For the rest of the analysis we proceed with these transformed
equations \eqref{eq:renamedClassicalFcts} keeping always in mind their
relation with the $(\xi,\eta)$ variables.

\begin{theorem}\label{thm:defbifdia}
  Let $(Z_{1},Z_{2})\in\bR^{*}\times\bR^{*}$, then {the bifurcation
    set of \eqref{def:enmomMapping} equals}
\[
\cB\l(\cF\ri)\rstr_{E\geq0} =
\l\{
(E,K)\in\cL \mid
E\geq 0
\text{ and }
K_{+}(E)\leq K \leq K_{-}(E)
\ri\}.
\]
  Here
$
\cL := \cL_{0}\cup\cL_{-}^{1}\cup\cL_{-}^{2}\cup\cL_{-}^{3}
       \cup\cL_{+}^{2}\cup\cL_{+}^{3}{\subset \bR^2}
$
  with
\beq
\begin{array}{lcl}
\cL_{0} := \{E=0\}
, &\qquad& \cL_{-}^{1} := \{K=Z_{-}-E\}, \\
\cL_{+}^{2} := \{K=-Z_{+}-E\}, &\qquad&
  \cL_{-}^{2} := \{K=-Z_{-}-E\}, \\
\cL_{+}^{3} := \{4EK=Z_{+}^{2}\}, &\qquad&
  \cL_{-}^{3} := \{4EK=Z_{-}^{2}\},
\end{array}
\Leq{eq:bifCurves}
  and $K_+$ and $K_-$ are defined by \eqref{eq:kplus}
  and \eqref{eq:kminus}.
\end{theorem}
\proof The fact that $K_{+}(E)\leq K \leq K_{-}(E)$ is a consequence
of Remark \ref{rem:boundsbif}.

$\{E = 0\}$ is the threshold between compact and non compact energy
surfaces. So by Definition \ref{def:bs} it belongs to the bifurcation
set $\cB\l(\cF\ri)$.

By definition, the critical points of $\cF$ are in
$\cB\l(\cF\ri)\rstr_{E\geq0}$.  To compute them we can take advantage
of the simple form of the level set equation in the $(x,y)$
coordinates.  To cover the plane we need to consider two half strips
(see Remark \ref{rmk:coord}.1).  We start assuming to cover the upper
half plane.  We can rewrite the level set equation
\[
\cF(p_{x},p_{y},x,y) =
\bsm
H(p_{x},p_{y},x,y) \\
H_{x}(p_{x},x)
\esm
=
\bsm
E \\
K
\esm
\]
in the form
\beq
\bem{l}
f_{1}(p_{y},y)\\
f_{2}(p_{x},x)
\eem
:=
\bem{l}
K + \frac{(1-y^{2})p_{y}^{2}}{2} + Z_{-} y + E y^{2} \\
K - \frac{(x^{2}-1)p_{x}^{2}}{2} + Z_{+} x + E x^{2}
\eem
=
\bem{l}
0 \\
0
\eem
\Leq{f1f2}
and use of this last representation to compute the critical points.

We look for values of $(p_{x}, p_{y}, x, y)$ such that
\[
D \bsm
f_{1}(p_{y},y)\\
f_{2}(p_{x},x)
\esm
=
\bsm
-(x^{2}-1)p_{x} & 0& -xp_{x}^{2} + Z_{+}+ 2E x & 0 \\
0 & (1-y^{2})p_{y} & 0 & -y p_{y}^{2} + 2 Ey +Z_{-}
\esm
\]
has rank smaller than $2$. Then, using $(x,y)\in
[1,\infty)\times[-1,1]$, it a simple exercise to check that the
critical points are given by the values
\begin{compactenum}
\item $(p_{x}, \pm\sqrt{Z_{-}+ 2E}, x,  1)$ and
$(p_{x}, \pm\sqrt{Z_{-}- 2E}, x, - 1)$,
\item $(p_{x}, 0, x, -Z_{-}/2E)$,
\item $(\pm\sqrt{Z_{+}+2E}, p_{y}, 1, y)$,
\item $(0, p_{y}, -Z_{+}/2E, y)$.
\end{compactenum}
Substituting these values in the equations \eqref{f1f2}, one obtains
the following curves in the $(E,K)$ plane
\begin{compactenum}
\item $K \pm Z_{-} + E = 0$,
  thus $\cL_{-}^{1,2}$ is in  $\cB\l(\cF\ri)$,
\item $K - \frac{Z_{-}^{2}}{4E} = 0$,
  thus $\cL_{-}^{3}$ is in  $\cB\l(\cF\ri)$,
\item $K + Z_{+} + E = 0$,
  thus $\cL_{+}^{2}$ is in $\cB\l(\cF\ri)$,
\item $K - \frac{Z_{+}^{2}}{4E} = 0$,
  thus $\cL_{+}^{3}$ is in $\cB\l(\cF\ri)$.
\end{compactenum}

For what concerns the lower half plane covered by points $(x,y)$, it
is enough to notice that its phase space equals the image of the one
considered under the inversion
$(p_x,p_{y},x,y)\mapsto(p_x,-p_{y},x,-y)$.  Therefore it reduces to
the analysis that we already performed.

We want to show that for all energy parameters $(E,K)$ in a connected
component of $\cR\setminus\cB(\cF)$ of the image $\cR$ of $\cF$, the
energy levels $\cF^{-1}(E,K)$ are diffeomorphic.  We start by
discussing a special example.

\begin{figure}[h!]
\begin{center}
\def\svgwidth{\linewidth}
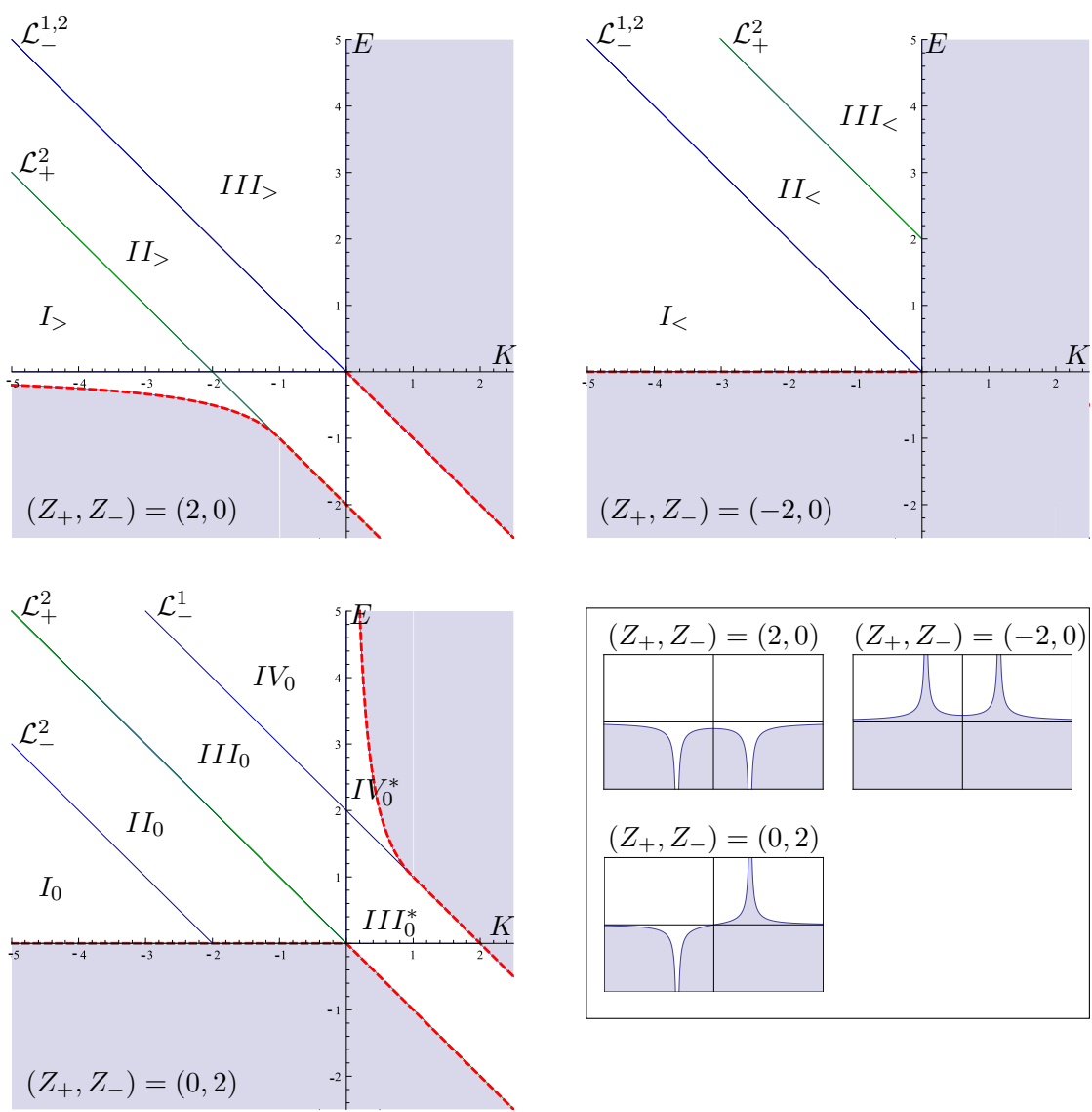
\caption{\small Examples of bifurcation diagrams for the considered
  planar cases, the shaded regions identify the complement of the
  Hill's region, the red dashed curves are $K_{+}$ and $K_{-}$. The
  green line $\cL_{+}^{2}$ parametrised by $K_0(E)=-(Z_++E)$
  corresponds to the closed orbit wandering between the two centers
  and having coordinate $x=1$. The box on the right shows the shape of
  the potential $V$ on the $q_{1}$-axis in the different cases.}
\label{fig:bif0}
\end{center}
\end{figure}

Let $(Z_+,Z_-) = (0,2)$. Let $(E_0,K_0)$ be in the interior of the
region bounded by $\cL_-^1$ and $K_-(E)$ (see Figure \ref{fig:bif0},
bottom left plot). We will show in Section \ref{sec:zp0} that all the
trajectories in configuration space with energy $(E_0,K_0)$ must cross
a segment $S_0$ strictly contained in the segment joining the two
centers (see Figure \ref{fig:motion02}, plot 5 and 6 counted from the
left). Since $(E_0,K_0)\not\in\cL$ the crossing must be
transversal. Therefore by the linearisability of the vector field we
can define a Poincar\'e section $S_0$, such that every trajectory is
uniquely identified by its crossing point (see \cite[Satz 3.46 and
Definition 7.16]{Kn11}).

Let $(E_1,K_1)$ be another point in the interior of the region
containing $(E_0,K_0)$. As before there is a segment $S_1$ strictly
contained in the segment joining the two centers that is crossed
transversally by all the trajectories (with energy $(E_1,K_1)$). Given
$(E_1,K_1)$ we can define the Poincar\'e section $S_1$ and every point
on the level set is identified by its crossing point and its time,
thus the level set is diffeomorphic to $\bR\times S_1$.

Clearly, $S_0$ and $S_1$ are diffeomorphic and thus the level sets of
$(E_0,K_0)$ and $(E_1,K_1)$ are diffeomorphic. By the generality of
$(E_0,K_0)$ and $(E_1,K_1)$ all the points in the interior of the
region bounded by $\cL_-^1$ and $K_-(E)$ have diffeomorphic level
sets.

We consider another example, again $(Z_+,Z_-) = (0,2)$. Let
$(E_0,K_0)$ be in the interior of the region of $\{E > 0\}$ bounded by
$\cL_+^2$, $\cL_-^1$ (see Figure \ref{fig:bif0}, bottom left plot). We
continue to refer to Section \ref{sec:zp0} when we show that for such
$(E,K)$ all the trajectories in configuration space must cross a line
segment $L_0$ strictly contained in the $q_1$-axis with $q_1<-1$ (see
Figure \ref{fig:coordPE} and the first plot from the right in Figure
\ref{fig:motion02}).

As in the previous example the trajectories must cross $L_0$
transversally and we can reduce the phase space to the Poincar\'e
section $L_0$. If $(E_1,K_1)$ is another point in the same region we
can reiterate the procedure to find a Poincar\'e section $L_1$ that is
diffeomorphic to $L_0$. And thus all the points in the region have
diffeomorphic level sets.

The argument sketched above can be reproduced in each connected
component of $\cR\setminus\cB(\cF)$ choosing a proper transversal
section. How to make the choice will be clear in the next three
sections, where we characterise the motion in configuration space for
the energy parameters in each region.  \qed

\begin{remarks}\label{rmk:redundance}
\begin{compactenum}
\item Differently from $\cL_0 = \{E=0\}$, the line $\{K=0\}$ is in the
  bifurcation set only in the symmetric case $Z_-=0$. In this case, in
  fact, it corresponds to the boundary $K_-(E)$ of the Hill's region.

  Of course there may be points $\{K=0\}$ in the bifurcation set for
  $Z_-\neq0$, but these are just the points in which curves in $\cL$
  cross $\{K=0\}$ transversally (see Figures \ref{fig:bif1} and
  \ref{fig:bif2}).
\item The characterisation of the bifurcation set given in Theorem
  \ref{thm:defbifdia} is redundant. Namely some of the curves
  $\cL_{*}^{*}$ restricted to the values of $(K,E)$ in the Hill's
  region could be empty for some values of $Z_{1}$ and $Z_{2}$.

  For example, being $K \leq K_-(E)$, we can immediately see that the
  curve $\cL_+^3$ will be in the bifurcation diagram for positive $E$
  only when $Z_+ < 0$ and $|Z_+| < Z_-$.  \hfill$\Diamond$
\end{compactenum}
\end{remarks}

In what follows we will describe more precisely the structure of the
bifurcation sets and of the trajectories in configuration space in
relation to the values assumed by $Z_{+}$ and $Z_{-}$.

The momenta $(p_{x}, p_{y})$ at given $(E,K)$ are given
in general by
\beq
p_{x}^{2} = \frac{2(x^{2}-1)(E x^{2} + Z_{+} x + K)}{(x^{2}-1)^{2}},
\quad
p_{y}^{2} = \frac{-2(1-y^{2})(E y^{2} +Z_{-} y + K)}{(1-y^{2})^{2}}.
\Leq{eq:momenta}

The Hill's region is identified by the values of $E$ and $K$ that
admit non-negative squared momenta.  The denominators in
\eqref{eq:momenta} being always positive, we can discuss them and
identify the possible motion types in terms of the zeros of the
numerators, that is, the polynomials \beq P_\pm(s) := 2 (s^{2}-1) (E
s^{2} + Z_{\pm} s + K).  \Leq{eq:critPoly} Here $s\in\{x,y\}$ and the
understanding is that we choose ``$+$'' for $s=x$ and ``$-$'' for
$s=y$. The factor $(s^{2}-1)$ is introduced to provide the correct
signs and for computational convenience. The momenta can be simply
obtained via $(x^{2}-1)p_{x} = \pm\sqrt{P_+(x)}$ and $(1-y^{2})p_{y} =
\pm\sqrt{P_-(y)}$.  The roots of $P_{+}(x)$ and $P_{-}(y)$ are
respectively
\begin{align*}\eqnumtag\label{eq:solsbif}
 x_{1,2} &= \pm1, \qquad \textstyle x_{3,4} =
   -\frac{Z_{+}}{2E} \pm \sqrt{\frac{Z_{+}^{2}}{4E^{2}}-\frac KE}, \\
 y_{1,2} &= \pm1, \qquad \textstyle y_{3,4} =
   -\frac{Z_{-}}{2E} \pm \sqrt{\frac{Z_{-}^{2}}{4E^{2}}-\frac KE},
\end{align*}
with the convention that the smaller index corresponds to the solution
with negative sign. In both variables, the polynomials have two fixed
roots at $\pm1$ and two movable roots which depend on the constants of
motion. Being $x\in[1,\infty)$, we are going to consider only roots in
this region.

The discriminant of $P_\pm$ is proportional to
\beq
\discr(P_\pm) = (Z_{\pm}^{2}-4EK) (E+K-Z_{\pm})^{2}(E+K+Z_{\pm})^{2}.
\Leq{eq:discr}
Double roots appear when $\discr(P_\pm)$ vanishes. For each couple
$(Z_{+},Z_{-})$ this gives six curves in the $(K,E)$-plane, three for
the $x$ variable and three for the $y$. These are the curves
$\cL_{+}^{1} := \{K=Z_{+}-E\}$ and $\cL_{-}^{1}$, $\cL_{\pm}^{2,3}$
defined by \eqref{eq:bifCurves}.

\begin{remark}
  The zeroes of the discriminant $P_\pm$ \eqref{eq:discr} correspond
  to the double roots of $(E s^{2} + Z_{\pm} s + K)$ and the points in
  which these roots reach the fixed roots $\pm 1$, i.e. the
  $q_{1}$-line. The positivity of $P_\pm$ and the positions of its
  roots, as we will see, characterise the trajectories in
  configuration space.

  The curve $\cL_+^1$ appearing in the discriminant depends from the
  fact that we considered $x\in\bR$. As such it will have no
  correspondence in the bifurcation set or in the description of the
  possible motions.
\hfill$\Diamond$\end{remark}

As a first step we consider the cases in which $Z_{-}=0$ or
$Z_{+}=0$. In these cases the $(K,E)$-plane is divided by the curves
$\cL_{\pm}^{1,2,3}$ into different regions.  We will label these
regions using roman numbers with a subscript chosen between $>$, $<$
and $0$ indicating if $Z_{+}>0$, $Z_{+}<0$ or $Z_{+}=0$
respectively. In Figure \ref{fig:bif0} are shown representative
bifurcation diagrams for these three cases with the corresponding
enumeration of the regions.

\subsection{Motion for $Z_{-} = 0$}\label{sec:zm_0}
The case $Z_{-} = 0$ corresponds to two attracting (or repelling)
centers with the same relative strength. We have the following
corollary of Theorem \ref{thm:defbifdia}.

\begin{corollary}
  Let $Z_{-} = 0$ and $Z_{+}\in\bR^{*}$. With the notation of
  \eqref{eq:bifCurves} we have
\[
\cB(\cF)\rst{E\geq0} = \l\{ (E,K)\in\cL_{0}\cup\cL_{-}^{1}\cup\cL_{+}^{2}
  \mid E\geq0 \mbox{ and } K_{+}(E)\leq K \leq 0\ri\}.
\]
\end{corollary}
\proof
In \eqref{1} we assumed $Z_1,Z_2 \neq 0$, therefore $Z_- = 0$
implies $Z_+ \neq 0$.

$K$ cannot be positive because $K_{-}(E) = 0$ for $E\geq0$.  By
$\cL_{-}^{1} = \cL_{-}^{2}$ and $\cL_{-}^{3} = \cL_{0}$, it is
redundant to add both in the definition of the bifurcation diagram.
The fact that $\cL_{+}^{3}$ is not in the bifurcation set follows from
Remark \ref{rmk:redundance}.2.  Then the claim follows directly from
Theorem \ref{thm:defbifdia}.  \qed

Consider the bifurcation diagram for $(Z_{+}, Z_{-}) = (\pm2,0)$
appearing in Figure \ref{fig:bif0}.  It is possible to describe the
qualitative structure of the motion in configuration space for energy
parameters in the different regions by studying the dynamic of the
roots \eqref{eq:solsbif} with respect to a walk in the bifurcation
diagram. By this we mean fixing a value of $E$ big enough and varying
$K$ to move through the regular regions $I_{>,<}$, $II_{>,<}$,
$III_{>,<}$ and to cross the bifurcation lines.

\begin{figure}[h]
\begin{center}
\makebox[\textwidth][c]{\includegraphics[width=1.\linewidth]{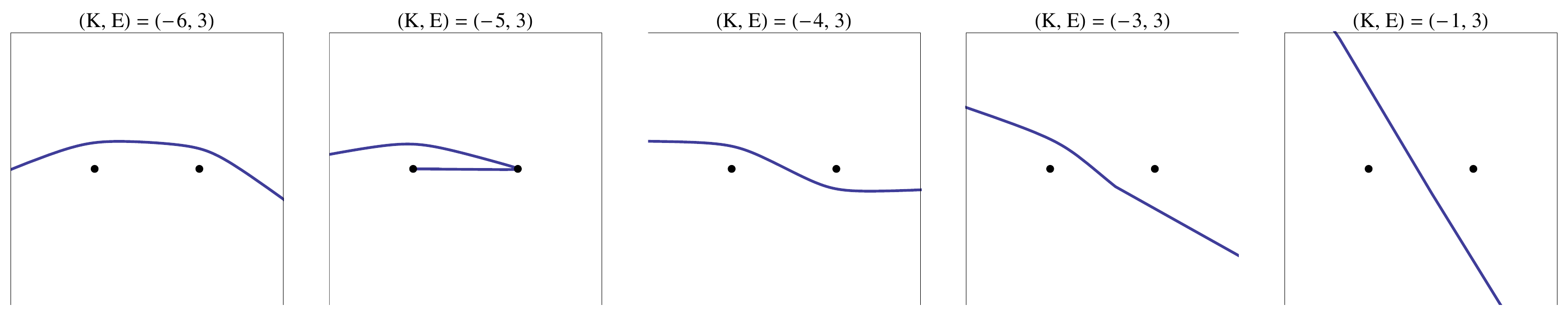}}
\caption{\small Example of possible trajectories in the case
  $Z_{+}=2$, $Z_{-}=0$ for $E=3$ and growing values of $K$ (from left
  to right) chosen in the different regions of the bifurcation
  diagram.}
\label{fig:motion20}
\end{center}
\end{figure}

\begin{figure}[h]
\begin{center}
\makebox[\textwidth][c]{\includegraphics[width=1.\linewidth]{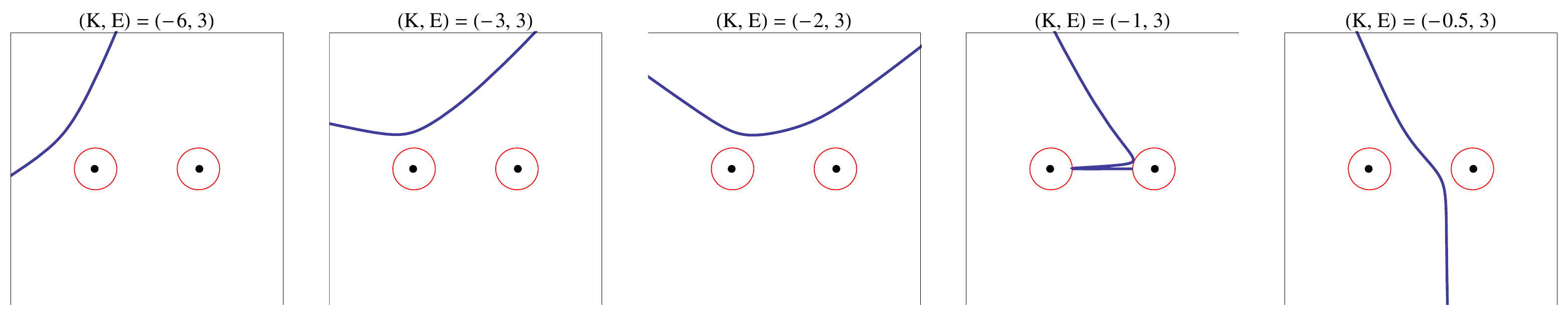}}
\caption{\small Example of possible trajectories in the case
  $Z_{+}=-2$, $Z_{-}=0$ for $E=3$ and growing values of $K$ (from left
  to right) chosen in the different regions of the bifurcation
  diagram. The red line corresponds to the energy level of the plotted
  trajectory.}
\label{fig:motion-20}
\end{center}
\end{figure}

A qualitative representation of how the motion changes with respect to
the energy parameters is shown in Figure \ref{fig:motion20} for $Z_{+}
= 2$ and in Figure \ref{fig:motion-20} for $Z_{+} =-2$. This can be
schematically explained through the behaviour of the roots
\eqref{eq:solsbif} as follows.

\begin{compactitem}
\item Roots of the polynomial $P_-(y)$.
\begin{compactitem}
\item For energies in the regions $I_{>,<}$ and $II_{>}$ of the
  bifurcation diagrams, the polynomial $P_-(y)$ is non-negative for
  every value of $y\in[-1,1]$ and $|y_{3,4}|>1$. Thus for energy
  parameters in these regions, the particle is allowed travel in
  configuration space everywhere in a region around the centers.
\item Line $\cL_{-}^{1}$ characterises the values $(E,K)$ such that
  the two groups of roots of $P_-(y)$ merge: $y_{3} =y_{1}=-1$ and
  $y_{4}=y_{2}=1$.
\item In the regions $II_{<}$ and $III_{>,<}$, $|y_{3,4}|<1$ and the
  $P_-(y)$ is not negative only if $y\in[y_{3},y_{4}]$. This means
  that the motion in configuration space can cross the axis through
  the two centers only passing through the segment between the
  centers.
\end{compactitem}
\item Roots of the polynomial $P_+(x)$.
\begin{compactitem}
\item Notice that $x_{1}$ and $x_{3}$ are always smaller than $1$,
  thus they do not belong to the domain of definition $[1,\infty)$ for
  $x$.
\item In the regions $I_{>,<}$ and $II_{<}$, the root $x_{4} >1$ and
  $P_+(x)$ is non negative only if $x\in[x_{4},\infty)$. Therefore in
  configuration space, the particle cannot reach the line joining the
  two centers.
\item For $(E,V)\in\cL^{2}_{+}$, we have the collision of the
  solutions $x_{4}=x_{2}=1$ and in configuration space the particle
  can reach the line between the centers.
\item On the right of $\cL^{2}_{+}$, in the regions $II_{>}$ and
  $III_{>/<}$, the root $x_{4} < 1$ and $P_+(x)$ is non-negative for
  $x\in[1,\infty)$. In other words the particle can cross the line in
  configuration space connecting the centers.
\end{compactitem}
\end{compactitem}

We can now understand the peculiarity of the lines in the bifurcation
set.
\begin{compactitem}
\item For values of the parameters on the singular line $\cL_{-}^{1}$
  we can identify two special trajectories in configuration space. In
  these, the particle lies in the positive (negative) $q_{1}$-axis
  with $|q_{1}|>1$, possibly bouncing against the singularity and
  being reflected back.
\item For $(E,V)$ on $\cL^{2}_{+}$ we can find the unique periodic
  orbit of the regularised classical two-centers problem, see
  \cite{Kn}. It is the hyperbolic trajectory of a particle bouncing
  between the centers. Counting from left to right, the second plot of
  Figure \ref{fig:motion20} and the fourth plot of Figure
  \ref{fig:motion-20} show the trajectory of a particle on the stable
  manifold of this special orbit.
\item For $K=0$ ($E>0$) only one trajectory is possible: the vertical
  trajectory moving on the line $y=0$.
\end{compactitem}

\begin{remark}
  Notice that the ordering of the singular curves reflects the main
  difference between the cases $Z_{+} > 0$ and $Z_{+} < 0$. In the
  first case (corresponding to the attracting potential) the particle
  is able to travel arbitrarily near to the centers. In the case
  $Z_{+} < 0$ the centers $\pm a$ have a positive distance from the
  Hill's region.
\hfill$\Diamond$\end{remark}

\subsection{Motion for $Z_{+} = 0$}\label{sec:zp0}

\begin{corollary}
Let $Z_{+} = 0$ and $Z_{-}>0$. With the notation of
\eqref{eq:bifCurves} we have
\[
\cB(\cF)\rst{E\geq0} = \Big\{ (E,K)\in{}
\cL_{0}\cup\cL_{-}^{1}\cup\cL_{-}^{2}\cup\cL_{-}^{3}\cup\cL_{+}^{2}
\mid  E\geq0 \mbox{ and } K_{+}(E)\leq K \leq K_{-}(E)\Big\}.
\]
\end{corollary}
\proof In \eqref{1} we assumed $Z_1,Z_2 \neq 0$, therefore $Z_+ = 0$
implies $Z_- \neq 0$.

We have $\cL_{+}^{3} = \cL_{0}$. The claim follows directly from
Theorem \ref{thm:defbifdia}.
\qed

\begin{figure}[ht]
\begin{center}
\makebox[\textwidth][c]{\includegraphics[width=1.\linewidth]{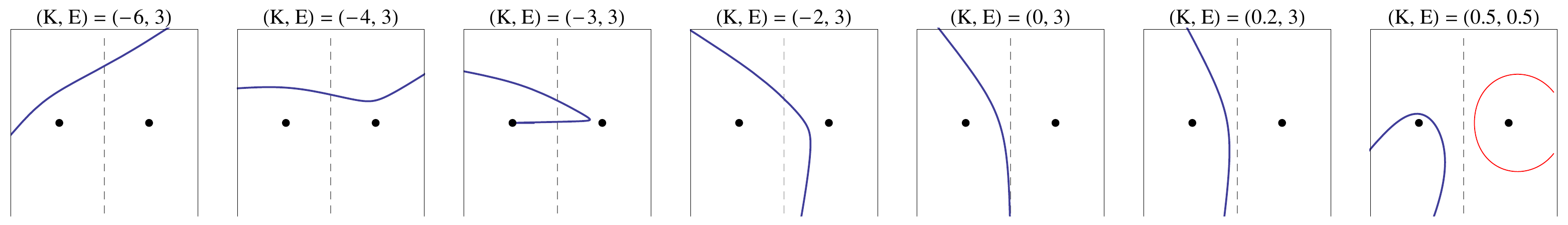}}
\caption{\small Example of possible motions in the case $Z_{+}=0$,
  $Z_{-}=2$ for values of $E$ and $K$ in different regions of the
  bifurcation diagram.  The red line is the energy level of the
  trajectory in the plot.}
\label{fig:motion02}
\end{center}
\end{figure}

As in the previous section we give a qualitative explanation of the
possible motions in configuration space through the behaviour of the
roots \eqref{eq:solsbif}. A visual support is provided by Figure
\ref{fig:motion02}.

\begin{compactitem}
\item Roots of the polynomial $P_-(y)$.
\begin{compactitem}
\item For $(E,V)$ in $I_{0}$, the polynomial $P_-(y)$ is not negative
  for any $y\in[-1,1]$. Thus in configuration space the particle is
  free to move around the centers.
\item For energy parameters on $\cL_{-}^{2}$, two roots collide:
  $y_{4}=y_{2}=1$.
\item The motion in configuration space for $(E,V)$ in $II_{0}$ and
  $III_{0}$ is restricted to $y\in[-1,y_{4}]$. I.e. the particle is
  free to travel around the attracting center but is bounded away from
  the repelling one.
\item For energies on $\cL_{-}^{1}$, the other two roots collide:
  $y_{3} = y_{1}=-1$ and for $(E,V)\in IV_{-}$ the only allowed $y$
  are restricted in $y\in[y_{3}, y_{4}]$: in configuration space the
  particle cannot anymore travel around the centers.
\item On the line $K=0$, $y_{4} = 0$ and for bigger values of $K$
  (i.e. in the regions $III_{0}^{*}$ and $IV_{0}^{*}$) $y_{4}$ becomes
  negative. The particle in configuration space is no more able to
  flow around the repelling center.
\end{compactitem}
\item For the roots of $P_+(x)$ the discussion is similar as before.
\begin{compactitem}
\item The roots $x_{1,3}$ are negative. We consider only the roots
  $x_{2,4}$.
\item For energy parameters in $I_{0}$ and $II_{0}$ the root $x_{4} >
  1$ and the polynomial $P_+(x)$ is non-negative for
  $x\in[x_{4},\infty)$. Therefore in configuration space the particle
  cannot reach the line between the centers.
\item For energy parameters on the right of $\cL_{+}^{2}$ the motion
  becomes possible for $x\in[1,\infty)$. I.e. the particle can reach
  the line between the centers.
\end{compactitem}
\end{compactitem}

\subsection{Motion in the general case}\label{bif:GenCase}
\begin{figure}[h!!]
\begin{center}
\def\svgwidth{1.\linewidth}
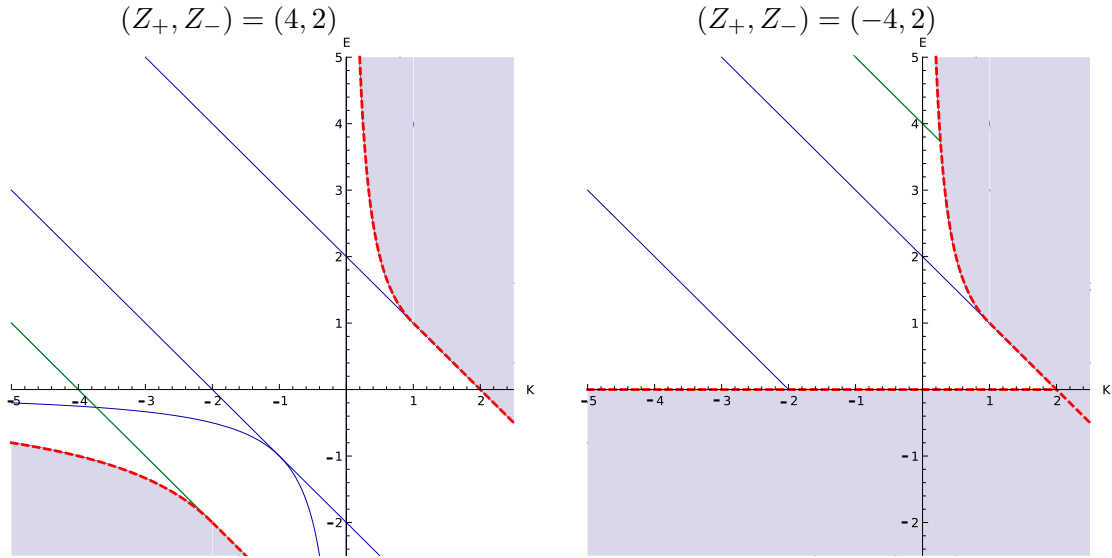
\caption{\small Bifurcation diagrams for the fully attracting (left)
  and the fully repelling (right) case respectively. The shaded
  regions identify the complement of the Hill's region.}
\label{fig:bif1}
\end{center}
\end{figure}

We first describe the bifurcation set for the fully repelling (or
attracting) configuration $\sign(Z_{1}) = \sign(Z_{2})$.  The picture
is similar to the one with $Z_{-} = 0$ (see Section \ref{sec:zm_0})
with the only difference that some positive values of $K$ are allowed
(see Figure \ref{fig:bif1}).

\begin{corollary}
  Let $|Z_{+}| > Z_{-}$, $Z_+\in\bR^*$, $Z_{-}\in\bR^{0}_{+}$. With
  the notation of \eqref{eq:bifCurves} we have
\[
\cB(\cF)\rst{E\geq0} =
\Big\{ (E,K)\in\cL_{0}\cup\cL_{-}^{1}\cup\cL_{-}^{2}\cup\cL_{-}^{3}\cup\cL_{+}^{2}
\mid  E\geq0, K_{+}(E)\leq K \leq K_{-}(E)\Big\}.
\]
\end{corollary}
\proof By Remark \ref{rmk:redundance}.2, $\cL_{+}^{3}$ is not in the
bifurcation set.  The corollary follows immediately from Theorem
\ref{thm:defbifdia}.
\qed

The structure of the trajectories and the qualitative behaviour of the
motion in configuration space for this case is analogous to the one
presented in Section \ref{sec:zm_0}, therefore we will not discuss it.

\begin{figure}[h!!]
\begin{center}\makebox[\textwidth][c]{
\def\svgwidth{1.\linewidth}
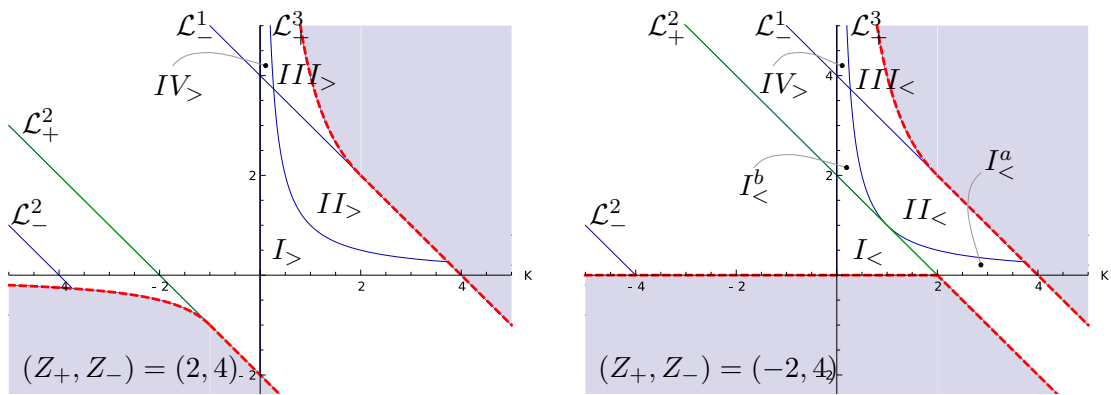}
\caption{Bifurcation diagrams and labeled regions for $|Z_{+}| <
  Z_{-}$. The shaded regions identify the complement of the Hill's
  region.}
\label{fig:bif2}
\end{center}
\end{figure}

The case $|Z_{+}| < Z_{-}$ (that is, $\sign(Z_{1}) \neq \sign(Z_{2})$)
is particularly interesting, since for positive energies a set of
bounded orbits of positive Liouville measure arises.

\begin{figure}[h!!]
\begin{center}
\makebox[\textwidth][c]{\includegraphics[width=1.\linewidth]{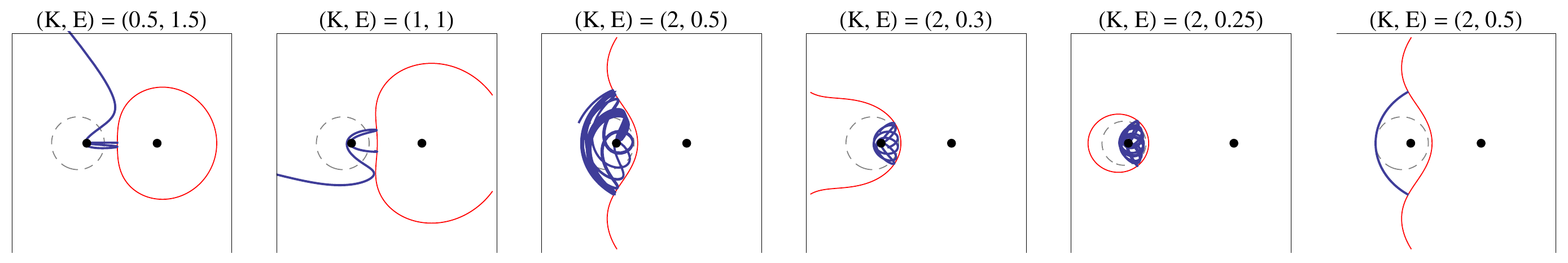}}
\caption{\small Bounded motions for $Z_+ < 0$, $|Z_{+}|< Z_{-}$ and
  $E\geq 0$. From left to right: trajectory for energy parameters on
  the boundary $\cL^{2}_{+}$ of $I^{a}_{<}$, trajectory for energies
  on the tangency point between $\cL^{2}_{+}$ and $\cL_{+}^{3}$,
  trajectory for energies on the boundary $\cL_{+}^{3}$ of $I^{a}_{<}$
  Then follow two trajectories for energies inside $I_{<}^{a}$ and the
  trajectory on $\cL^{3}_{+}$. The red line is the energy level of the
  trajectory in the plot, the dotted line is the $0$-energy level.}
\label{fig:bOrbit}
\end{center}
\end{figure}

\begin{corollary}\label{thm:GenBifDia}
  Let $|Z_{+}| < Z_{-}$, $Z_+\in\bR^*$, $Z_{-}\in\bR_{+}$. If $Z_{+}
  >0$ we have
\[
\cB(\cF)\rst{E\geq0} =
\Big\{ (E,K)\in \cL_{0}\cup\cL_{-}^{1}\cup\cL_{-}^{2}\cup\cL_{-}^{3}
\cup\cL_{+}^{2} \mid E\geq0 \mbox{ and }
  K_{+}(E)\leq K \leq K_{-}(E)\Big\},
\]
while if $Z_{+} < 0$ we have
\[
\cB(\cF)\rst{E\geq0} =
\Big\{ (E,K) \in \cL_{0}\cup\cL_{-}^{1}\cup\cL_{-}^{2}\cup\cL_{-}^{3}
\cup\cL_{+}^{2}\cup\cL_{+}^{3} \mid E\geq0 \mbox{ and }
  K_{+}(E)\leq K \leq K_{-}(E)\Big\}.
\]
\end{corollary}
\proof By Remark \ref{rmk:redundance}.2, $\cL_{+}^{3}$ is in the
bifurcation set for $Z_+<0$.  The corollary follows immediately from
Theorem \ref{thm:defbifdia}.
\qed

In this case the qualitative behaviour of the motion in configuration
space is analogous to the one presented in Section \ref{sec:zp0}, but
for energy parameters in the region $I_{<}^{a}$.  This region contains
the set of energy parameters included in the region bounded above by
$\cL_{+}^{3-}:=\l\{ (E,K)\in\cL_{+}^{3} \Mid E <
\frac{|Z_{+}|}{2}\ri\}$, on the sides by $\cL_{+}^{2}$ and
$\cL_{+}^{1}$ and below by $E\geq 0$ (see Fig.\ \ref{fig:bif2}).

For $(E,V)$ on $I_{<}^{a}$ a new phenomenon appears: both the movable
roots of $P_+(x)$ are bigger than one and the polynomial is
non-negative in the union of the two disjoint intervals $[1,x_{3}]$
and $[x_{4},\infty)$.  On the configuration space they give rise to an
escaping trajectory similar to the previous ones and to a family of
bounded trajectories near the attracting center (see Figure
\ref{fig:bOrbit}).

\section*{Acknowledgements}
The author wants to express his gratitude to Andreas Knauf and Mirko
Degli Esposti for their help, the helpful discussions and their many
useful comments.  The author acknowledges partial support by the EPSRC
grant EP/J016829/1 and the FIRB-project RBFR08UH60 (MIUR, Italy).

\addcontentsline{toc}{section}{References}

\bibliographystyle{amsalpha}
\bibliography{biblio}
\end{document}